\documentclass{article}

\usepackage{amssymb,amsmath}
\usepackage{epsfig,multirow}

\usepackage{subfigure}

\newcommand\x{{\bf x}}

\newcommand\zero{{\bf 0}}

\begin{document}
\title{Tracking Many Solution Paths of a Polynomial Homotopy \\
       on a Graphics Processing Unit\thanks{This 
  material is based upon work supported by the National
  Science Foundation under Grant ACI-1440534.}
}
\author{Jan Verschelde\thanks{email: {\tt jan@math.uic.edu},
        URL: {\tt www.math.uic.edu/}$\sim${\tt jan}}
    ~and~ Xiangcheng Yu\thanks{email: {\tt xyu30@uic.edu}} \\
Department of Mathematics, Statistics, and Computer Science \\
University of Illinois at Chicago, 851 South Morgan (M/C 249) \\
Chicago, IL 60607-7045, USA}

\date{1 May 2015}

\maketitle

\begin{abstract}
\noindent Polynomial systems occur in many areas of science and engineering.
Unlike general nonlinear systems, the algebraic structure enables
to compute all solutions of a polynomial system.
We describe our massive parallel predictor-corrector algorithms
to track many solution paths of a polynomial homotopy.
The data parallelism that provides the speedups stems from the
evaluation and differentiation of the monomials in the same polynomial
system at different data points, which are the points on the solution paths.
Polynomial homotopies that have tens of thousands of solution paths 
can keep a sufficiently large amount of threads occupied.
Our accelerated code combines the reverse mode of algorithmic differentiation
with double double and quad double precision to compute more accurate
results faster.

\noindent {\bf Keywords.}
algorithmic differentiation,
continuation methods, 
double double,
Graphics Processing Unit (GPU),
path tracking, polynomial system,
polynomial homotopy,
quad double.
\end{abstract}

\section{Introduction}

Many problems in computational algebraic geometry can be solved by 
computing solutions of polynomial systems.  As the number of solutions
can grow exponentially in the degrees, the number of variables and
equations, the computational complexity of these problems are hard.
GPUs provide a promising technology to deliver
significant speedups over traditional processors, but may require
a complete overhaul of the algorithms we use
in our polynomial system solvers.

This paper describes the application of numerical continuation methods
to track many solution paths defined by a polynomial homotopy.
A polynomial homotopy is a family of polynomial systems in which
the solutions depend on one real parameter~$t$.  Starting at $t = 0$,
there are many solution paths originating at known solutions of a
start system.  These paths end at $t = 1$, 
at solutions of a polynomial system we want to solve.
A common homotopy links the start system $g(\x) = \zero$
to the target system $f(\x) = \zero$ linearly as
\begin{equation} \label{eqlinhom}
   \gamma (1-t) g(\x) + t f(\x) = \zero, 
\end{equation}
where $\gamma$ is a random complex constant.
The random $\gamma$ ensures that paths originating at the regular solutions 
of $g(\x) = \zero$ will stay regular for all $t < 1$.
For this regularity result to hold, all arithmetic must be complex.
Our problem is
the tracking of $m$ solution paths in complex $n$-space.

The difficulty with implementing the predictor-corrector algorithms
to track solution paths defined by polynomial homotopies is that
the traditional formulation of the algorithms does not match the
data parallelism for which GPUs are designed for.
For instance, while each solution path can be tracked independently,
the number of predictor-corrector stages may fluctuate significantly
depending on the geometric shape of the path.  The type of high level
parallelism that is applied in distributed memory message passing
or shared memory multithreaded processors does not apply to our problem.
On GPUs, a relatively small number of multiprocessors
can independently launch a large number of threads that perform the same
synchronized sequences of instructions.

Applying Newton's method as a corrector, in every step,
we evaluate all polynomials and all their derivatives in the system.
To achieve a high level of parallelism for this task, the terms in
each polynomial are decomposed as the product of the variables that
occur in the term with a positive exponent and the factor that is
common in all derivatives of the term.  
The reverse mode of algorithmic differentiation~\cite{GW08} applied 
to each product of variables then reaches a high level of parallelism.
Its cost matches optimal complexity bounds for the evaluation 
and differentiation problem.
The linear systems in each Newton step we solve in the least squares
sense via the modified Gram-Schmidt method.
In~\cite{VY12} and~\cite{VY13} our computations were executed on
randomly generated regular data sets.  In~\cite{VY14}, we integrated
and improved the evaluation and differentiation codes to run
Newton's method on some selected benchmark polynomial systems.
We extended this in~\cite{VY15} to track one {\em single} path
of a polynomial homotopy on a GPU.
The focus in this paper is on the tracking of {\em many} paths.

For applications, achieving speedup is not enough, the numerical results
must be accurate as well.  As the degrees and the number of solution paths
increase, the numerical conditioning is likely to worsen as well.
To improve the quality, we calculate with double double and
quad double arithmetic, using the QD library~\cite{HLB00} on the host
and its CUDA version~\cite{LHL10} on the device.
While for complex double arithmetic, the evaluation and differentiation
algorithms are memory bound, for complex double double and quad double
arithmetic, these algorithms become compute bound.
The double digit speedups compensate for the extra cost overhead caused
by the extended precision.  With the accelerated versions of our code
we are often able to obtain more accurate results faster than without
the use of GPUs.  We obtain speedup and {\em quality up}.

Although solving polynomial systems may seem a very specific problem
(we refer to~\cite{Li03} for a survey),
many software packages have been developed for this problem, e.g.:
Bertini~\cite{Bertini}, 
HOM4PS~\cite{GLL02}, HOM4PS-2.0~\cite{LLT08}, HOM4PS-3~\cite{CLL14},
PHoM~\cite{GKKTFM04}, NAG4M2~\cite{Ley11},
and HOMPACK~\cite{WBM87,WSMMW97}.
Many of these packages are still under active development.
To the best of our knowledge, our code provides the first path tracker
for polynomial systems for a GPU.

Related research in computer algebra concerns the implementation of
polynomial operations on GPUs.
Reports on this research are~\cite{HLMMPX14} and~\cite{MP11}.  
Computer algebra is geared towards exact computations,
often over finite number fields.  Our approach is numerical
and we improve the accuracy of our results with double double
and quad double arithmetic.  This type of arithmetic is described
in the section of error-free transformations in~\cite{Rum10}.
Interval arithmetic on CUDA GPUs~\cite{CDD12} is an alternative approach
to improve the quality of numerical computations.

The description of our path tracker consists of two parts.
First we define the scheduling of the threads
and then we outline the organization of the memory.
Polynomial systems with ten of thousands of solutions
are at the bottom of the threshold for which we start
to notice the beneficial effect of our accelerated codes.

\newpage
\section{SIMT Path Tracking}

The Single Instruction Multiple Threads (SIMT) execution model in GPUs
implies that threads are either executing the same instruction or they
are idle.  In our application, the threads are evaluating and differentiating
the same polynomial system, at different approximations for the solutions 
along the path.
In the SIMT model, we run the same arithmetic circuit at different data.
Table~\ref{tabsimtprecor} shows a simplified model.

\begin{table}[hbt]
\begin{center}
\begin{tabular}{c|c|c}
path0 & path1 & path2 \\ \hline
{\tt predict} & {\tt predict} & {\tt predict} \\ \hline
{\tt evaluate} & {\tt evaluate} & {\tt evaluate} \\
{\tt correct} & {\tt correct} & {\tt correct} \\ \hline
{\tt evaluate} & & {\tt evaluate} \\
{\tt correct} &  & {\tt correct} \\ \hline
              &  &  {\tt evaluate}\\
              &  & {\tt correct}  \\
\end{tabular}
\caption{Simplified SIMT predictor-corrector steps on three paths.
The first path needs three, the second path needs only one,
and the third path needs two {\tt evaluate}-{\tt correct} steps
to converge.}
\label{tabsimtprecor}
\end{center}
\end{table}

We distinguish three stages in a predictor-corrector algorithm.
The first stage is called {\tt predict} in Table~\ref{tabsimtprecor}.
The predictor consists of the application of an extrapolation algorithm,
applied to each coordinate of the solution separately.
Typically, a fourth order predictor uses the current and four previous
points on the path to predict the coordinates of the solution.
This stage is naturally parallel and has a linear cost in the dimension.
In the {\tt evaluate} stage, the polynomial system is evaluated and
differentiated at the solution.
As this stage can have a cost that is cubic (or higher) in the dimension,
it is separate from the {\tt correct} stage.  In the {\tt correct} stage,
a linear system is solved to compute the update to the solution.

For every path we execute a prediction step and at least one
{\tt evaluate}-{\tt correct} step.
Some paths may need two or even
three such steps for Newton's method to converge so the residuals
and size of the updates are sufficiently small.
While the instructions are the same, it is important that all the data
are distinct.  Not only the coordinates of each solution, 
but also the value for the continuation parameter~$t$ 
and the step size differ.
The length of the total execution time is bounded from below by the
time required for the most difficult solution path.

For memory considerations, paths that have been tracked to their end
are relabeled and their work space in memory is swapped to the end.
In the schematic of Table~\ref{tabsimtprecor2}, the paths of
Table~\ref{tabsimtprecor} are reordered.

\begin{table}[hbt]
\begin{center}
\begin{tabular}{c|c|c}
job0 & job1 & job2 \\ \hline
{\tt predict0} & {\tt predict1} & {\tt predict2} \\ \hline
{\tt evaluate0} & {\tt evaluate1} & {\tt evaluate2} \\
{\tt correct0} & {\tt correct1} & {\tt correct2} \\ \hline
{\tt evaluate0} & {\tt evaluate2} & \\
{\tt correct0} &  {\tt correct2} &  \\ \hline
{\tt evaluate2} & &  \\
{\tt correct2} &  &  \\
\end{tabular}
\caption{Simplified SIMT predictor-corrector steps on three paths.
For each stage, each job is associated with its $path\_idx$ and empty jobs are removed.}
\label{tabsimtprecor2}
\end{center}
\end{table}

When we track one single path, as in~\cite{VY15},
the step size control can be performed by the host.
When tracking many solution paths, every solution path has its own step size
and current value of the continuation parameter~$t$.
In this situation, the step size control is executed on the device.

After each evaluation and correction, there is a check kernel to determine the success 
status is represented as 0, $-1$ and 1. 0 is to continue, $-1$ is failure and 1 is success. 
A parallel scan~\cite{SHG08} to count of all paths with 0's can generate the $path\_idx$ 
for the new round, see Table~\ref{tabjobidx}.

\begin{table}[hbt]
\begin{center}
\begin{tabular}{c|cccccc}
             & path0 & path1 & path2 & path3 & path4 &$\cdots$ \\ \hline
status       & $0$   & $1$   & $0$   & $-1$  & $0$ & $\cdots$ \\ 
scan for $0$ & $1$   & $1$   & $2$   & $2$   & $3$ & $\cdots$ \\
$job\_idx$   & $1$   &       & $2$   &       & $3$ & $\cdots$ \\
$path\_idx$  & $0$   &       & $2$   &       & $4$ & $\cdots$ 
\end{tabular}
\caption{Generated $path\_idx$ from current iteration status for the next round of computation}
\label{tabjobidx}
\end{center}
\end{table}

Thus, the only number passed between CPU and GPU each step is total number of paths to continue,
which can be easily determine from the the last element of scan.

\section{SIMT Evaluation and Differentiation}

The speedups we obtain are mainly due to the fine granularity of
the arithmetic circuits to evaluate and differentiate the polynomials
in the system.  Table~\ref{tabcpugpu} outlines the major differences
in the organization of the algorithms for the host (CPU) and the device (GPU).

\begin{table}[hbt]
\caption{At the top is pseudo code for the host to evaluate and 
differentiate a polynomial system.
At the bottom is the corresponding pseudo code for the device. }
\begin{center}
\begin{tabbing}
\hspace{1cm} \=
   {\em Pseudo code on the host:} \\
\> for\= ~each polynomial do \\
\>    \> for\= ~each monomial do \\
\>    \>    \> 1. \= compute the coefficient $c(t)$ for this monomial; \\
\>    \>    \> 2. evaluate the monomial and its derivative; \\
\>    \>    \> 3. add the values to the polynomials 
                  and to the Jacobian matrix. \\
\\
\> {\em Pseudo code on the device:} \\
\>    \> lau\=nch the following three kernels \\
\>    \> 1. \= compute the coefficient $c(t)$ \\
\>    \>    \> for all monomials in all polynomials; \\
\>    \> 2. evaluate the monomial and its derivatives \\
\>    \>    \> for all monomials in all polynomials; \\
\>    \> 3. add to the value of the polynomial and to the Jacobian matrix \\
\>    \>    \> for all monomials in all polynomials.
\end{tabbing}
\label{tabcpugpu}
\end{center}
\end{table}

For evaluation of single path, the evaluated and differentated monomials are the operands in a long summation operation, executed to calculate the evaluated polynomials and the evaluated Jacobian matrix of the system.
The values of the evaluated and differentiated monomials are positioned 
in an irregular pattern in memory.  Therefore to the sum kernel, it
appears as if the data is at random positions in memory. But for evaluations of multiple paths, all terms, at the same postion of Jacobian matrix, follow the same instruction to sum the same positions of its own path. If the matrix of evaluated and differentiated monomials is transposed, the sum kernel can benefit from memory coalescing.
The transposition is illustrated in Table~\ref{tabtranspose}.

\begin{table}[hbt]
\begin{center}
\begin{tabular}{c}
\begin{tabular}{c|cccc}
       & \multicolumn{4}{c}{monomials in memory} \\ \hline
path 0 & $a_0 a_1 a_2$ & $a_1 a_2$ & $a_0 a_2$ & $a_1 a_2$ \\
path 1 & $b_0 b_1 b_2$ & $b_1 b_2$ & $b_0 b_2$ & $b_1 b_2$ \\
path 2 & $c_0 c_1 c_2$ & $c_1 c_2$ & $c_0 c_2$ & $c_1 c_2$ \\
$\cdots$ & $\cdots$ & $\cdots$ & $\cdots$ & $\cdots$
\end{tabular} \\
\quad $\Downarrow$ \quad \\
\begin{tabular}{c|c|c|c}
path 0 & path 1 & path 2 & $\cdots$ \\ \hline
$a_0 a_1 a_2$ & $b_0 b_1 b_2$ & $c_0 c_1 c_2$ & $\cdots$ \\ 
$a_1 a_2$ & $b_1 b_2$ & $c_1 c_2$ & $\cdots$ \\ 
$a_0 a_2$ & $b_0 b_2$ & $c_0 c_2$ & $\cdots$ \\ 
$a_1 a_2$ & $b_1 b_2$ & $c_1 c_2$ & $\cdots$ \\
\end{tabular}
\end{tabular}
\caption{Tranposition of the matrix of evaluated and differentiated
monomials, evaluated at different points.  In the example,
we consider $x_0 x_1 x_2$ at the points
$(a_0, a_1, a_2)$, $(b_0, b_1, b_2)$, $(c_0, c_1, c_2)$, $\ldots$.}
\label{tabtranspose}
\end{center}
\end{table}

Instead of transposing matrix after monomial evaluation, we redesign the monomial kernel. The reverse mode~\cite{GW08} is used to generate vertical values and the same monomial of multiple paths are computed together in blocks. This directly fit our goal to transpose the monomial values. Also, all threads in each block shared the same instructions to evaluate monomials, which save the instruction reading time. The organization of the evaluation of a monomial and its
derivatives is displayed in Table~\ref{tabmoneval}.

\begin{table}[hbt]
\begin{center}
\begin{tabular}{c|c|c|c}
& \multicolumn{3}{c}{$x_1 x_2 x_3 x_4$ and its four
                   derivatives evaluated} \\ \cline{2-4}
& path 0 & path 1 & path 2 \\ \hline
0 & $a_1$ & $b_1$ & $c_1$ \\
1 & $a_1 \star a_2$ & $b_1 \star b_2$ & $c_1 \star c_2$ \\
2 & $a_1 a_2 \star a_3$ & $b_1 b_2 \star b_3$ & $c_1 c_2 \star c_3$ \\
7 & $a_1 a_2 a_3 \star a_4$
  & $b_1 b_2 b_3 \star b_4$ & $c_1 c_2 c_3 \star c_4$ \\
6 & $a_1 a_2 \star a_4$ & $b_1 b_2 \star b_4$ & $c_1 c_2 \star c_4$ \\
3 & $a_3 \star a_4$ & $b_3 \star b_4$ & $c_3 \star c_4$ \\
4 & $a_1 \star a_3 a_4$ & $b_1 \star b_3 b_4$ & $c_1 \star c_3 c_4$ \\
5 & $a_2 \star a_3 a_4$ & $b_2 \star b_3 b_4$ & $c_2 \star c_3 c_4$ \\
\end{tabular}
\caption{The sequence of steps in evaluating one monomial
and its derivatives for three paths at different points
$(a_1, a_2, a_3, a_4)$, $(b_1, b_2, b_3, b_4)$, and
$(c_1, c_2, c_3, c_4)$.
Each new multiplication is marked by a $\star$. }
\label{tabmoneval}
\end{center}
\end{table}

Compared with the tree mode~\cite{VY14}, 
this consecutive mode has more memory bandwidth. Although monomial evaluation part has twice memory 
access than tree mode, summation has more speedup due to consecutive memory. 
Also, multiple threads in a single block use the same instruction to avoid redundant reading. 
Thus, this consecutive mode is more suitable for evaluation of multiple paths.

\begin{table}[hbt]
\begin{center}
\begin{tabular}{c|c|rrr} 
   & name & double & double double & quad double \\ \hline
\multirow{3}{*}{Mon} &{\tt cyclic10} & 190.41 &  124.78 & 25.70 \\
&{\tt nash8}    & 206.68 &  143.30 & 27.62 \\
&{\tt pieri44}  & 209.47 &  147.31 & 27.32 \\ \hline
\multirow{3}{*}{Sum} &{\tt cyclic10} & 104.91 &  126.63 & 123.13 \\
&{\tt nash8}    & 121.38 &  128.52 & 126.56 \\
&{\tt pieri44}  & 87.26  &  80.41  & 77.56
\end{tabular}
\caption{Memory bandwidth of 1,000 evaluations of the same polynomial system in complex(GB/s)}
\label{tabbench}
\end{center}
\end{table}

All path join its work space of evaluation vertically, and a relatively small matrix transpose of Jacobian is used before correction. Linear systems in the corrector are solved with a $QR$ decomposition.
Table~\ref{tabmemory} shows the organization of the memory.

\begin{table}[hbt]
\begin{center}
\begin{tabular}{|c|c||c|c|c|c|} \hline
\multicolumn{2}{|c||}{instructions} 
& \multicolumn{4}{|c|}{work space for multiple path} \\ \hline \hline
idx & cff
& \multicolumn{3}{|c|}{evaluate-joint} & correct \\ \hline
\multicolumn{2}{c}{~}
& \multicolumn{1}{|c|}{cff} & mon & Jac & Jac', $R, \Delta \x$ \\ \cline{3-6}
\end{tabular}
\caption{Schematic organization of the memory.  Instructions to
evaluate and differentiate a polynomial system are stored by
indices (idx) and coefficients (cff).
We store the coefficients (cff), evaluated at the value
of the continuation parameter~$t$, the evaluated monomials (mon),
and the Jacobian matrix (Jac) joint vertically for all path. For corrector, the transpose of Jac,  $R$ and the update $\Delta \x$ to the solution.}
\label{tabmemory}
\end{center}
\end{table}

\section{Computational Results}

We developed and executed our code on a Linux workstation.
Our benchmark applications were selected for the diversity
of the research areas and their size.
Because our application benefit the most from the accelerated
evaluation and differentiation algorithms, we report first on
computations done separately from the path tracking.

\subsection{Hardware and Software}

Our code is compiled with version 6.5 of the CUDA Toolkit
and {\tt gcc -O2}.
A Red Hat Enterprise Linux workstation of Microway,
with Intel Xeon E5-2670 processors at 2.6~GHz is the
host for the NVIDIA Tesla K20C,
which has 2496 cores with a clock speed of 706~MHz.
To prepare the benchmark data we used Python, in particular
{\tt phcpy}~\cite{Ver14},
the Python interface to PHCpack~\cite{Ver99}.

The double double and quad double arithmetic
is provided by the QD library~\cite{HLB00}.
This QD library has been ported to GPUs,
we used the code available at~\cite{LHL10}.

\subsection{Applications}

We selected three examples of polynomial systems,
which arose in different applications.
The examples can be formulated for any number of equations and variables.
We selected three systems and in each case we applied
the homotopy~(\ref{eqlinhom}) to solve the systems.
Below is a brief description of each system:

(1) {\tt cyclic10:} the cyclic 10-roots problem is a 10-dimensional
system with 34,940 isolated complex solutions.  Except for the last
equation (which has two terms), every polynomial has 10 monomials.
The $k$-th polynomial in this system is of degree~$k$.
These roots appear in the study of complex Hadamard matrices~\cite{Szo11}.

(2) {\tt nash8}: the solutions of this system give all totally mixed 
Nash equilibria in a game with 8 players, where each player has two
pure strategies, see~\cite{Dat09}, \cite{MM97}.
For generic payoff matrices, this 8-dimensional system has 14,833 equilibria.
Every polynomial in this system has 130 monomials of degrees ranging from one
till seven.

(3) {\tt pieri44}: there are 24,024 four dimensional planes that 
meet 16 four dimensional planes, given in general position.
This system is a 16-dimensional problem and can be interpreted as
a matrix completion problem~\cite{KRW04}, see also~\cite{HSS98}, \cite{HV00}.
Every polynomial in the system is of degree 4 and has 246 monomials.

As is typical for polynomial systems, 
the number of isolated solutions grows exponentially in
the dimensions and the degrees.
The systems we selected are large enough to notice a benefit of
the accelerated code and small enough so we can still compute all
isolated solutions.
Table~\ref{tabsumchar} summarizes their characteristics.

\begin{table}[hbt]
\begin{center}
\begin{tabular}{c|rrr} 
    name       & dim & \#paths & \#monomials \\ \hline
{\tt cyclic10} & 10~ &  34,940 &    92~~~ \\
{\tt nash8}    &  8~ &  14,833 & 1,040~~~ \\
{\tt pieri44}  & 16~ &  24,024 & 3,936~~~
\end{tabular}
\caption{Name, dimension, number of paths, and number of monomials
in the benchmark applications.}
\label{tabsumchar}
\end{center}
\end{table}

The number of isolated solutions equals the number of
solution paths in the polynomial homotopy.
In a massively parallel application we launch about 10,000
of threads.  In these applications, the parallelism comes
from evaluating the same system at about 10,000
different solution paths.

\subsection{Evaluations}

In Tables~\ref{tabevalcyclic10}, \ref{tabevalnash8}, 
and~\ref{tabevalpieri44} we list times and speedups for the
evaluation of the three systems.
With the number of simultaneous evaluations we go as far as 
the memory of the device allows us.
The speedups were computed by taking the time on the K20C
over the time on one 2.6~GHz CPU.
With the NVIDIA profiler, times on the GPU are reported for the
three kernels (defined in Table~\ref{tabcpugpu}),
in the columns with headers mon, sum, and coeff.

Although the number of variables is small (10, 8, and 16),
there are already sufficiently many monomials to achieve
a large enough parallelism to obtain good speedups.
For double arithmetic, the problem is memory bound.
The speedups become really good in complex double double
and quad double arithmetic, because then the problem
is compute bound.

\begin{table}[hbt]
\begin{center}
{\small
\begin{tabular}{r|r|rrrr|r}
\multicolumn{7}{c}{complex double arithmetic} \\
     &  CPU  & \multicolumn{4}{c|}{GPU} &      \\
\#evals &  total & mon & sum & coeff & total & speedup \\ \hline
  10 &  0.062 &  0.017 &  0.008 &  0.004 &  0.028 &  2.19 \\
  20 &  0.078 &  0.020 &  0.008 &  0.004 &  0.033 &  2.39 \\
  50 &  0.188 &  0.024 &  0.011 &  0.005 &  0.040 &  4.69 \\
 100 &  0.379 &  0.030 &  0.016 &  0.006 &  0.051 &  7.39 \\
 150 &  0.553 &  0.038 &  0.021 &  0.007 &  0.066 &  8.41 \\
 200 &  0.732 &  0.042 &  0.026 &  0.008 &  0.076 &  9.60 \\
 250 &  0.928 &  0.049 &  0.032 &  0.009 &  0.090 & 10.31 \\
 300 &  1.132 &  0.056 &  0.037 &  0.010 &  0.103 & 11.04 \\
 500 &  1.824 &  0.087 &  0.056 &  0.015 &  0.157 & 11.61 \\
 750 &  2.786 &  0.126 &  0.079 &  0.021 &  0.226 & 12.32 \\
1000 &  3.748 &  0.155 &  0.101 &  0.026 &  0.282 & 13.30 \\
1250 &  4.748 &  0.203 &  0.127 &  0.032 &  0.363 & 13.08 \\
1500 &  5.563 &  0.235 &  0.149 &  0.039 &  0.423 & 13.14 \\
2000 &  7.381 &  0.299 &  0.191 &  0.050 &  0.540 & 13.67 \\
3000 & 11.148 &  0.459 &  0.284 &  0.082 &  0.826 & 13.50 \\
\multicolumn{7}{c}{complex double double arithmetic} \\
     &  CPU  & \multicolumn{4}{c|}{GPU} &      \\
\#evals &  total & mon & sum & coeff & total & speedup \\ \hline
  10 &   0.587 &  0.066 &  0.011 &  0.011 &  0.088 &  6.65 \\
  20 &   1.135 &  0.066 &  0.012 &  0.011 &  0.089 & 12.79 \\
  50 &   2.808 &  0.072 &  0.017 &  0.012 &  0.101 & 27.90 \\
 100 &   5.598 &  0.092 &  0.028 &  0.017 &  0.137 & 40.81 \\
 150 &   8.601 &  0.121 &  0.036 &  0.022 &  0.179 & 48.03 \\
 200 &  11.225 &  0.145 &  0.043 &  0.025 &  0.213 & 52.64 \\
 250 &  13.951 &  0.154 &  0.053 &  0.029 &  0.236 & 59.11 \\
 300 &  16.821 &  0.181 &  0.060 &  0.037 &  0.278 & 60.56 \\
 500 &  27.912 &  0.263 &  0.092 &  0.052 &  0.408 & 68.47 \\
 750 &  41.877 &  0.379 &  0.137 &  0.074 &  0.590 & 71.01 \\
1000 &  55.871 &  0.472 &  0.175 &  0.096 &  0.743 & 75.24 \\
1250 &  69.835 &  0.587 &  0.220 &  0.117 &  0.924 & 75.54 \\
1500 &  83.920 &  0.691 &  0.257 &  0.139 &  1.087 & 77.20 \\
2000 & 112.040 &  0.917 &  0.338 &  0.183 &  1.438 & 77.92 \\
3000 & 167.568 &  1.383 &  0.502 &  0.278 &  2.163 & 77.47 \\
\multicolumn{7}{c}{complex quad double arithmetic} \\
     &  CPU  & \multicolumn{4}{c|}{GPU} &      \\
\#evals &  total & mon & sum & coeff & total & speedup \\ \hline
  10 &    5.572 &  0.632 &  0.042 &  0.072 &  0.705 &   7.91 \\
  20 &   11.129 &  0.622 &  0.043 &  0.073 &  0.738 &  15.07 \\
  50 &   27.769 &  0.633 &  0.054 &  0.075 &  0.762 &  36.44 \\
 100 &   55.566 &  0.931 &  0.080 &  0.130 &  1.141 &  48.70 \\
 150 &   83.369 &  1.213 &  0.098 &  0.179 &  1.491 &  55.92 \\
 200 &  111.027 &  1.438 &  0.120 &  0.224 &  1.782 &  62.29 \\
 250 &  138.872 &  1.428 &  0.144 &  0.235 &  1.808 &  76.82 \\
 300 &  166.546 &  1.641 &  0.161 &  0.277 &  2.079 &  80.11 \\
 500 &  277.978 &  2.486 &  0.257 &  0.436 &  3.178 &  87.46 \\
 750 &  416.268 &  3.435 &  0.369 &  0.594 &  4.398 &  94.64 \\
1000 &  554.742 &  4.582 &  0.485 &  0.786 &  5.853 &  94.77 \\
1250 &  694.084 &  5.715 &  0.591 &  0.943 &  7.249 &  95.75 \\
1500 &  833.445 &  6.809 &  0.699 &  1.183 &  8.691 &  95.89 \\
2000 & 1111.412 &  8.916 &  0.929 &  1.532 & 11.377 &  97.69 \\
3000 & 1676.977 & 13.244 &  1.375 &  2.245 & 16.864 &  99.44 \\
\end{tabular}
}
\caption{Timings in milliseconds for the evaluation of the polynomials and
the Jacobian matrix of the cyclic 10-roots problem on the CPU and GPU.}
\label{tabevalcyclic10}
\end{center}
\end{table}

\begin{table}[hbt]
\begin{center}
{\small
\begin{tabular}{r|r|rrrr|r}
\multicolumn{7}{c}{complex double arithmetic} \\
     &  CPU  & \multicolumn{4}{c|}{GPU} &      \\
\#evals &  total & mon & sum & coeff & total & speedup \\ \hline
  10 &  0.311 &  0.042 &  0.050 &  0.015 &  0.106 &  2.92 \\
  20 &  0.586 &  0.057 &  0.069 &  0.015 &  0.072 &  8.10 \\
  50 &  1.417 &  0.079 &  0.075 &  0.027 &  0.181 &  7.81 \\
 100 &  2.813 &  0.140 &  0.113 &  0.032 &  0.285 &  9.86 \\
 150 &  4.181 &  0.202 &  0.134 &  0.045 &  0.380 & 11.00 \\
 200 &  5.586 &  0.244 &  0.169 &  0.057 &  0.470 & 11.89 \\
 250 &  6.927 &  0.295 &  0.187 &  0.064 &  0.546 & 12.68 \\
 300 &  8.302 &  0.349 &  0.224 &  0.078 &  0.651 & 12.75 \\
 500 & 13.834 &  0.567 &  0.314 &  0.125 &  1.006 & 13.75 \\
 750 & 20.650 &  0.863 &  0.470 &  0.193 &  1.527 & 13.53 \\
1000 & 27.509 &  1.111 &  0.608 &  0.254 &  1.973 & 13.94 \\
1250 & 34.433 &  1.435 &  0.772 &  0.319 &  2.526 & 13.63 \\
1500 & 41.253 &  1.682 &  0.892 &  0.390 &  2.964 & 13.92 \\
2000 & 55.157 &  2.209 &  1.179 &  0.523 &  3.910 & 14.11 \\
3000 & 82.710 &  3.303 &  1.742 &  0.877 &  5.922 & 13.97 \\
\multicolumn{7}{c}{complex double double arithmetic} \\
     &  CPU  & \multicolumn{4}{c|}{GPU} &      \\
\#evals &  total & mon & sum & coeff & total & speedup \\ \hline
  10 &   4.345 &  0.195 &  0.116 &  0.050 &  0.361 & 12.03 \\
  20 &   8.664 &  0.201 &  0.125 &  0.056 &  0.382 & 22.66 \\
  50 &  21.587 &  0.226 &  0.141 &  0.062 &  0.429 & 50.26 \\
 100 &  43.239 &  0.411 &  0.219 &  0.120 &  0.750 & 57.68 \\
 150 &  65.571 &  0.602 &  0.247 &  0.159 &  1.008 & 65.04 \\
 200 &  86.489 &  0.762 &  0.321 &  0.215 &  1.297 & 66.67 \\
 250 & 108.585 &  0.839 &  0.351 &  0.255 &  1.445 & 75.14 \\
 300 & 130.030 &  1.016 &  0.422 &  0.299 &  1.737 & 74.85 \\
 500 & 216.220 &  1.623 &  0.598 &  0.491 &  2.712 & 79.74 \\
 750 & 325.524 &  2.445 &  0.910 &  0.719 &  4.074 & 79.90 \\
1000 & 431.826 &  3.203 &  1.182 &  0.957 &  5.341 & 80.86 \\
1250 & 540.026 &  4.057 &  1.501 &  1.197 &  6.755 & 79.94 \\
1500 & 647.817 &  4.778 &  1.722 &  1.439 &  7.939 & 81.60 \\
2000 & 864.464 &  6.361 &  2.299 &  1.936 & 10.596 & 81.58 \\
3000 & 1301.577 &  9.517 &  3.420 &  2.984 & 15.922 & 81.75 \\
\multicolumn{7}{c}{complex quad double arithmetic} \\
        &  CPU  & \multicolumn{4}{c|}{GPU} &      \\
\#evals &  total & mon & sum & coeff & total & speedup \\ \hline
  10 &   43.425 &  1.956 &  0.502 &  0.506 &  2.964 &  14.65 \\
  20 &   86.566 &  1.977 &  0.522 &  0.534 &  3.033 &  28.55 \\
  50 &  216.214 &  2.154 &  0.552 &  0.537 &  3.244 &  66.66 \\
 100 &  433.039 &  4.150 &  0.807 &  1.051 &  6.009 &  72.07 \\
 150 &  650.377 &  5.990 &  0.878 &  1.569 &  8.437 &  77.09 \\
 200 &  866.149 &  8.051 &  1.171 &  2.077 & 11.299 &  76.66 \\
 250 & 1082.852 &  8.478 &  1.239 &  2.142 & 11.859 &  91.31 \\
 300 & 1297.308 & 10.046 &  1.474 &  2.537 & 14.057 &  92.29 \\
 500 & 2161.734 & 16.866 &  1.938 &  4.182 & 22.986 &  94.05 \\
 750 & 3245.388 & 25.008 &  2.874 &  6.189 & 34.071 &  95.25 \\
1000 & 4327.603 & 33.228 &  3.852 &  8.173 & 45.253 &  95.63 \\
1250 & 5407.600 & 41.280 &  4.769 & 10.153 & 56.202 &  96.22 \\
1500 & 6497.880 & 51.975 &  5.514 & 12.679 & 70.168 &  92.61 \\
2000 & 8652.404 & 68.903 &  7.380 & 16.727 & 93.010 &  93.03 \\
3000 & 12977.386 & 100.940 & 10.799 & 24.771 & 136.510 &  95.07 \\
\end{tabular}
}
\caption{Times in milliseconds for the evaluation of the polynomials and the
Jacobian matrix of the Nash equilibrium system on the CPU and the GPU.}
\label{tabevalnash8}
\end{center}
\end{table}

\begin{table}[hbt]
\begin{center}
\begin{tabular}{r|r|r|r}
\multicolumn{4}{c}{complex double arithmetic} \\
\#paths &  CPU  & GPU &  speedup \\
   10 &  0.152 &  0.196 &  0.77 \\
   20 &  0.330 &  0.239 &  1.38 \\
   50 &  0.815 &  0.292 &  2.79 \\
  100 &  1.512 &  0.341 &  4.43 \\
  200 &  2.894 &  0.462 &  6.26 \\
  500 &  7.257 &  0.809 &  8.97 \\
 1000 & 14.171 &  1.343 & 10.55 \\
 2000 & 28.524 &  2.514 & 11.35 \\
 5000 & 72.292 &  6.156 & 11.74 \\
\multicolumn{4}{c}{complex double double arithmetic} \\
\#paths &  CPU  & GPU &  speedup \\ \hline
   10 &   2.130 & 0.595 &  3.58 \\
   20 &   4.496 & 0.641 &  7.01 \\
   50 &  11.215 & 0.720 & 15.59 \\
  100 &  20.813 & 0.831 & 25.04 \\
  200 &  40.018 & 1.124 & 35.62 \\
  500 & 100.446 & 2.057 & 48.82 \\
 1000 & 194.243 & 3.462 & 56.11 \\
 2000 & 392.615 & 6.345 & 61.87 \\
 5000 & 992.708 & 15.504 & 64.03 \\
\multicolumn{4}{c}{complex quad double arithmetic} \\
\#paths &  CPU  & GPU &  speedup \\ \hline
  10 &   20.745 &  4.593 &   4.52 \\
  20 &   42.969 &  4.835 &   8.89 \\
  50 &  106.348 &  5.101 &  20.85 \\
 100 &  198.098 &  5.926 &  33.43 \\
 200 &  383.885 &  8.846 &  43.40 \\
 500 &  986.145 & 16.407 &  60.10 \\
1000 & 1876.226 & 28.365 &  66.15 \\
2000 & 3805.213 & 52.710 &  72.19 \\
5000 & 9618.930 & 128.948 &  74.60 \\
\end{tabular}
\caption{Times in seconds and speedups for tracking a number of paths
of the Nash equilibrium system.}
\label{tabpathnash8}
\end{center}
\end{table}

\begin{table}[hbt]
\begin{center}
{\small
\begin{tabular}{r|r|rrrr|r}
\multicolumn{7}{c}{complex double arithmetic} \\
        &  CPU  & \multicolumn{4}{c|}{GPU} &      \\
\#evals &  total & mon & sum & coeff & total & speedup \\ \hline
  10 &  1.129 &  0.137 &  0.138 &  0.049 &  0.324 &  3.48 \\
  20 &  2.127 &  0.168 &  0.156 &  0.050 &  0.373 &  5.70 \\
  50 &  5.223 &  0.239 &  0.208 &  0.097 &  0.544 &  9.60 \\
 100 & 10.226 &  0.447 &  0.306 &  0.113 &  0.866 & 11.80 \\
 150 & 15.303 &  0.654 &  0.396 &  0.162 &  1.212 & 12.63 \\
 200 & 20.239 &  0.794 &  0.475 &  0.206 &  1.475 & 13.72 \\
 250 & 25.369 &  0.971 &  0.575 &  0.237 &  1.783 & 14.23 \\
 300 & 30.387 &  1.157 &  0.671 &  0.289 &  2.116 & 14.36 \\
 500 & 50.778 &  1.890 &  1.113 &  0.471 &  3.474 & 14.62 \\
 750 & 75.665 &  2.895 &  1.589 &  0.729 &  5.213 & 14.51 \\
1000 & 102.170 &  3.718 &  2.074 &  0.958 &  6.751 & 15.13 \\
1250 & 127.156 &  4.821 &  2.698 &  1.215 &  8.735 & 14.56 \\
1500 & 154.357 &  5.645 &  3.126 &  1.494 & 10.265 & 15.04 \\
2000 & 201.537 &  7.425 &  4.064 &  2.003 & 13.492 & 14.94 \\
3000 & 302.158 & 11.108 &  6.138 &  3.351 & 20.597 & 14.67 \\
\multicolumn{7}{c}{complex double double arithmetic} \\
        &  CPU  & \multicolumn{4}{c|}{GPU} &      \\
\#evals &  total & mon & sum & coeff & total & speedup \\ \hline
  10 &  15.116 &  0.559 &  0.266 &  0.170 &  0.995 & 15.19 \\
  20 &  29.886 &  0.582 &  0.287 &  0.202 &  1.071 & 27.91 \\
  50 &  75.020 &  0.659 &  0.391 &  0.217 &  1.267 & 59.22 \\
 100 & 151.854 &  1.263 &  0.573 &  0.437 &  2.273 & 66.80 \\
 150 & 228.841 &  1.887 &  0.732 &  0.598 &  3.217 & 71.14 \\
 200 & 298.554 &  2.425 &  0.907 &  0.781 &  4.113 & 72.59 \\
 250 & 373.088 &  2.696 &  1.101 &  0.932 &  4.729 & 78.89 \\
 300 & 447.931 &  3.264 &  1.287 &  1.113 &  5.664 & 79.09 \\
 500 & 746.392 &  5.299 &  2.129 &  1.862 &  9.289 & 80.35 \\
 750 & 1120.713 &  8.042 &  3.104 &  2.726 & 13.873 & 80.79 \\
1000 & 1491.030 & 10.570 &  4.080 &  3.649 & 18.299 & 81.48 \\
1250 & 1867.582 & 13.400 &  5.217 &  4.568 & 23.185 & 80.55 \\
1500 & 2251.611 & 15.851 &  6.006 &  5.510 & 27.367 & 82.27 \\
2000 & 2990.387 & 21.057 &  7.908 &  7.429 & 36.394 & 82.17 \\
3000 & 4478.135 & 31.423 & 12.001 & 11.455 & 54.879 & 81.60 \\
\multicolumn{7}{c}{complex quad double arithmetic} \\
        &  CPU  & \multicolumn{4}{c|}{GPU} &      \\
\#evals &  total & mon & sum & coeff & total & speedup \\ \hline
  10 &  146.920 &  5.329 &  1.132 &  1.867 &  8.328 &  17.64 \\
  20 &  293.975 &  5.369 &  1.188 &  1.935 &  8.493 &  34.61 \\
  50 &  734.441 &  6.104 &  1.954 &  1.468 &  9.526 &  77.10 \\
 100 & 1470.332 & 12.760 &  2.123 &  3.895 & 18.778 &  78.30 \\
 150 & 2206.220 & 18.763 &  2.490 &  5.891 & 27.144 &  81.28 \\
 200 & 2942.909 & 25.181 &  3.149 &  7.859 & 36.189 &  81.32 \\
 250 & 3676.626 & 29.401 &  8.068 &  3.598 & 41.067 &  89.53 \\
 300 & 4415.752 & 30.987 &  4.207 &  9.617 & 44.810 &  98.54 \\
 500 & 7346.943 & 58.511 &  6.909 & 15.901 & 81.321 &  90.35 \\
 750 & 11049.214 & 86.933 &  9.707 & 23.621 & 120.261 &  91.88 \\
1000 & 14697.217 & 113.970 & 13.027 & 31.309 & 158.306 &  92.84 \\
1250 & 18394.761 & 134.100 & 16.487 & 38.865 & 189.452 &  97.09 \\
1500 & 22045.021 & 177.920 & 19.023 & 48.569 & 245.512 &  89.79 \\
\end{tabular}
}
\caption{Times in milliseconds for the evaluation of the polynomials and
the Jacobian matrix of the Pieri hypersurface system on the CPU and the GPU.}
\label{tabevalpieri44}
\end{center}
\end{table}

\begin{table}[hbt]
\begin{center}
\begin{tabular}{r|r|r|r}
\multicolumn{4}{c}{complex double arithmetic} \\
\#paths &  CPU  & GPU &  speedup \\
   10 &   0.757 &  0.506 &  1.50 \\
   20 &   1.580 &  0.603 &  2.62 \\
   50 &   3.883 &  0.890 &  4.36 \\
  100 &   7.800 &  1.229 &  6.35 \\
  200 &  15.813 &  1.801 &  8.78 \\
  500 &  39.861 &  3.713 & 10.74 \\
 1000 &  80.347 &  6.898 & 11.65 \\
 2000 & 161.498 & 13.232 & 12.21 \\
 5000 & 401.001 & 33.050 & 12.13 \\
\multicolumn{4}{c}{complex double double arithmetic} \\
\#paths &  CPU  & GPU &  speedup \\ \hline
   10 &   11.307 &  2.042 &  5.54 \\
   20 &   23.558 &  2.231 & 10.56 \\
   50 &   58.339 &  3.010 & 19.38 \\
  100 &  113.878 &  3.883 & 29.32 \\
  200 &  232.249 &  5.120 & 45.36 \\
  500 &  586.282 & 10.141 & 57.81 \\
 1000 & 1183.342 & 18.317 & 64.60 \\
 2000 & 2376.400 & 34.497 & 68.89 \\
\multicolumn{4}{c}{complex quad double arithmetic} \\
\#paths &  CPU  & GPU &  speedup \\ \hline
  10 &  111.498 & 19.403 &   5.75 \\
  20 &  234.984 & 20.642 &  11.38 \\
  50 &  583.908 & 25.590 &  22.82 \\
 100 & 1168.055 & 34.496 &  33.86 \\
 200 & 2375.275 & 47.696 &  49.80 \\
 500 & 5986.772 & 91.191 &  65.65 \\
1000 & 12075.740 & 165.244 &  73.08
\end{tabular}
\caption{Times in seconds and speedups for tracking a number of paths
of the hypersurface Pieri system.}
\label{tabpathpieri44}
\end{center}
\end{table}

\begin{figure}[hbt]
\begin{center}
\epsfig{figure=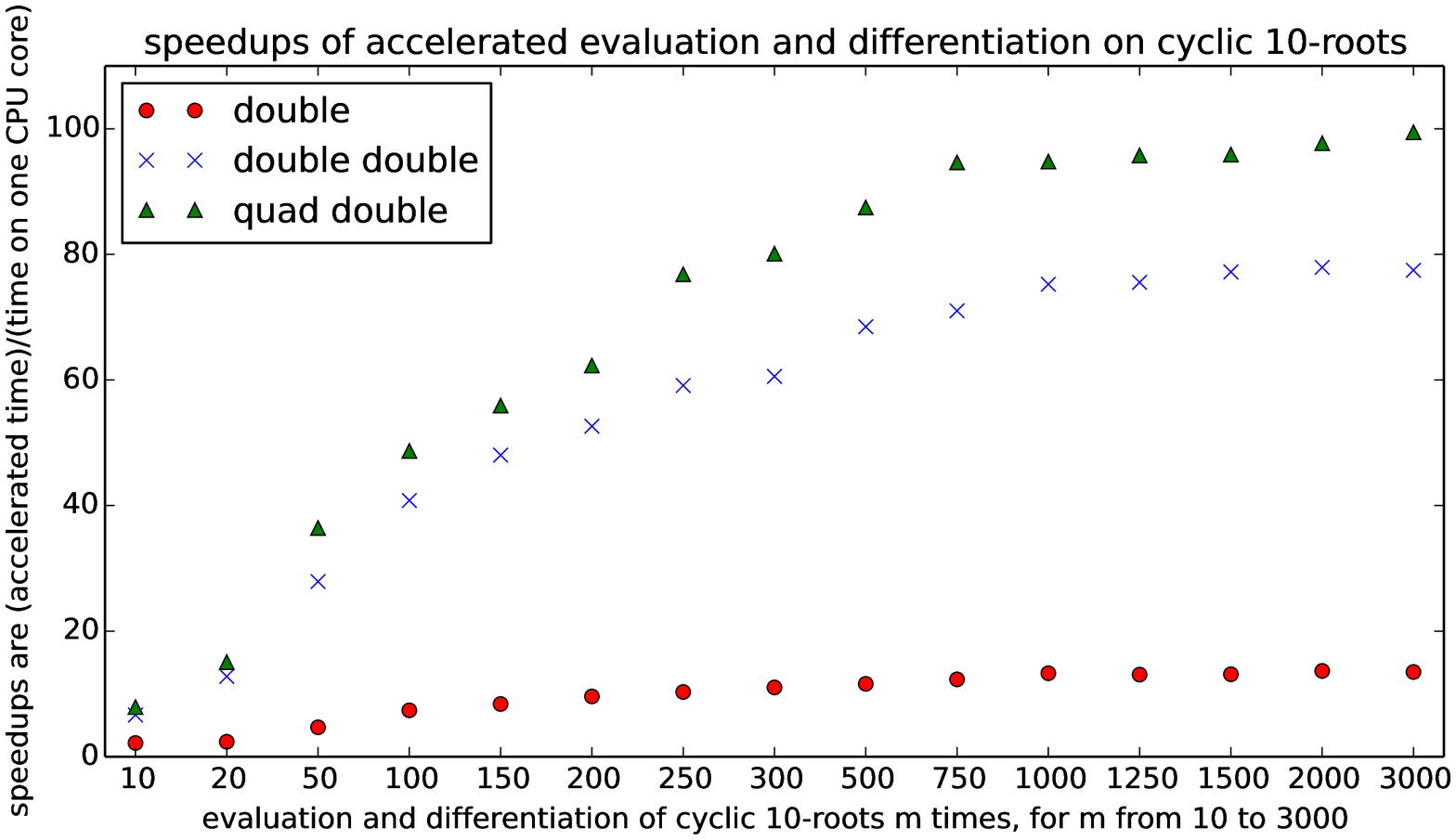, width=9.5cm}
\epsfig{figure=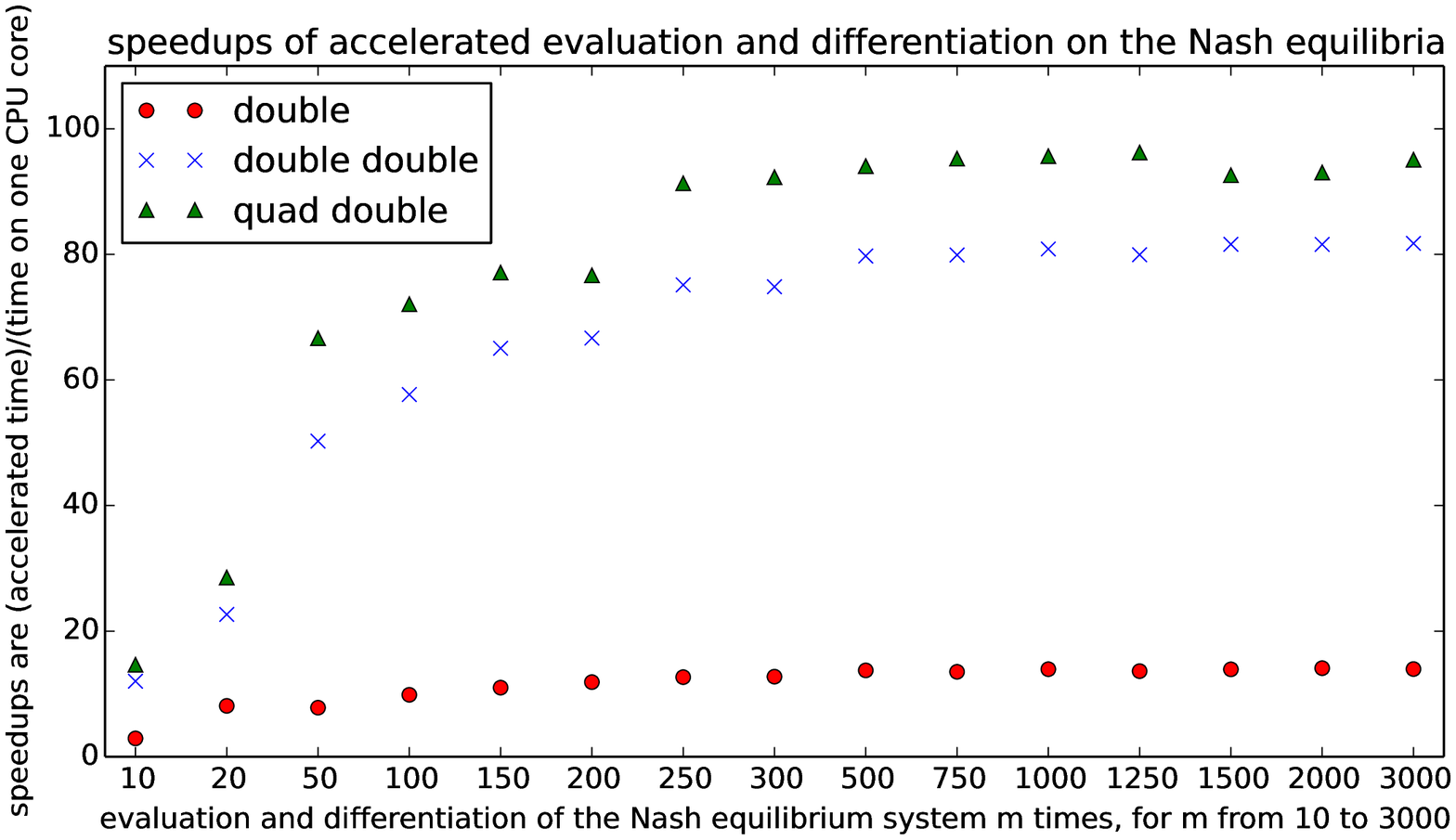, width=9.5cm}
\epsfig{figure=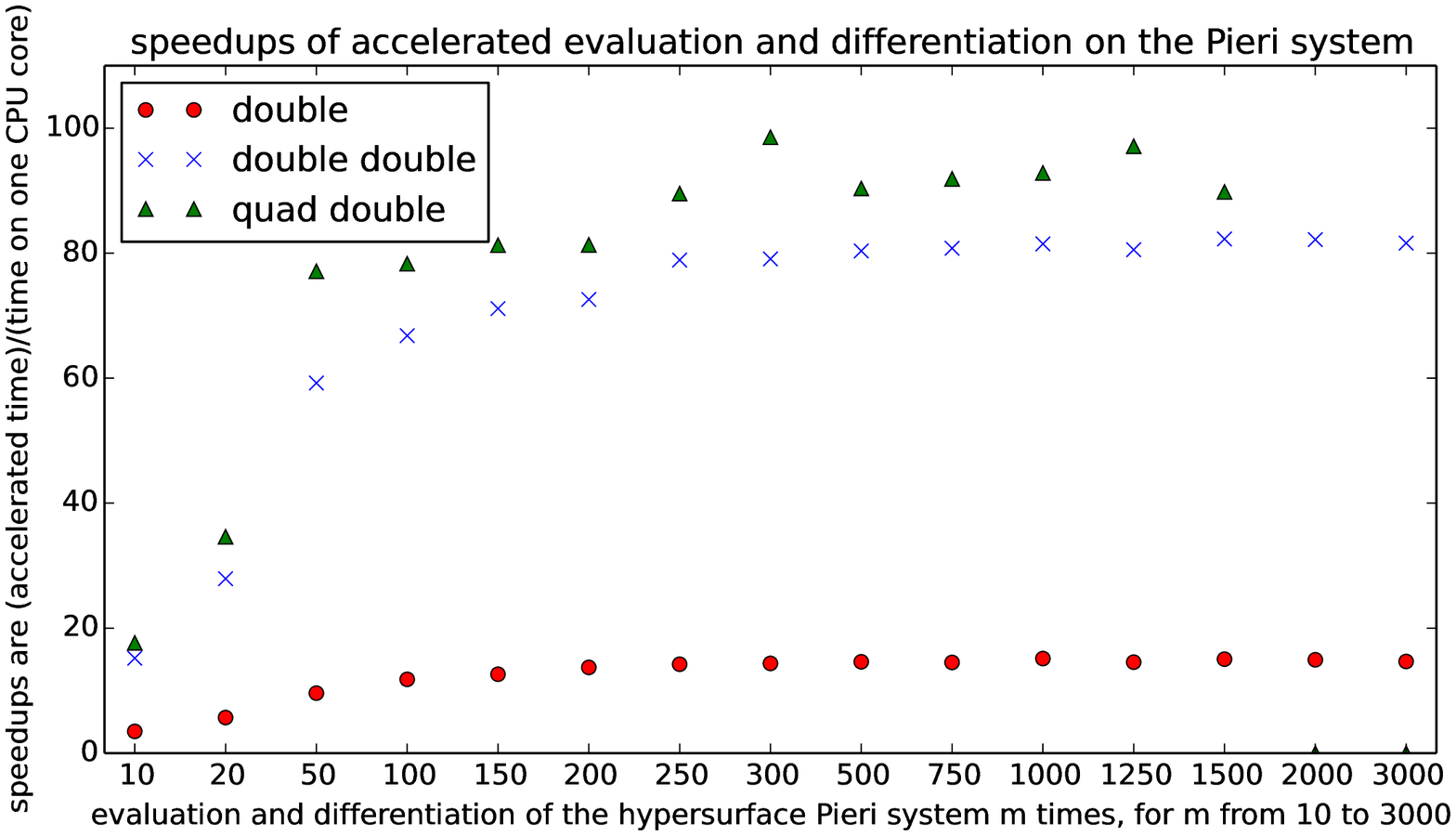, width=9.5cm}
\caption{Speedups for evaluating and differentiating the cyclic 10-roots
problem, the Nash equilibrium and the hypersurface Pieri systems 
in complex double, double double, and quad double arithmetic.}
\label{figeval}
% \label{figevalcyclic10}
\end{center}
\end{figure}

The corresponding speedups are shown in Figure~\ref{figeval}.
Observe that the more monomials the system has,
the fewer number of evaluations (respectively the fewer number of paths) 
are required to reach double digit speedups.

% \begin{figure}[hbt]
% \begin{center}
% \epsfig{figure=multevalnash8speedups.eps, width=9.5cm}
% \caption{Speedups for evaluating and differentiating the Nash equilibrium 
% system in complex double, double double, and quad double arithmetic.}
% \label{figevalnash8}
% \end{center}
% \end{figure}
% \begin{figure}[hbt]
% \begin{center}
% \epsfig{figure=multevalpieri44speedups.eps, width=9.5cm}
% \caption{Speedups for evaluating and differentiating the hypersurface
% Pieri problem in complex double, double double, and quad double arithmetic.}
% \label{figevalpieri44}
% \end{center}
% \end{figure}

\subsection{path tracking}

Results for tracking many paths for the cyclic 10-roots problem are
summarized in Table~\ref{tabpathcyclic10}.
Observe the quality up.  Tracking 10,000 paths in double double
arithmetic takes 10 seconds on the GPU, while on the CPU it takes
26.562 seconds in double arithmetic.
With our accelerated code we obtain solutions in a precision that
is twice as large in a time that is more than twice as fast.

\begin{table}[hbt]
\caption{Times in seconds and speedups for tracking a number of paths
of the cyclic 10-roots system.}
\begin{center}
\begin{tabular}{r|r|r|r}
\multicolumn{4}{c}{complex double arithmetic} \\
\#paths &  CPU  & GPU &  speedup \\
   10 &  0.040 &  0.128 &  0.31 \\
   20 &  0.075 &  0.139 &  0.54 \\
   50 &  0.158 &  0.147 &  1.07 \\
  100 &  0.277 &  0.155 &  1.79 \\
  200 &  0.482 &  0.181 &  2.67 \\
  500 &  1.239 &  0.250 &  4.96 \\
 1000 &  2.609 &  0.432 &  6.03 \\
 2000 &  5.341 &  0.768 &  6.96 \\
 5000 & 13.358 &  1.711 &  7.81 \\
10000 & 26.562 &  3.334 &  7.97 \\
\multicolumn{4}{c}{complex double double arithmetic} \\
\#paths &  CPU  & GPU &  speedup \\ \hline
   10 &   0.563 &  0.344 &  1.63 \\
   20 &   1.082 &  0.386 &  2.80 \\
   50 &   2.248 &  0.404 &  5.56 \\
  100 &   3.706 &  0.421 &  8.81 \\
  200 &   6.480 &  0.458 & 14.15 \\
  500 &  16.802 &  0.729 & 23.05 \\
 1000 &  35.683 &  1.315 & 27.14 \\
 2000 &  83.601 &  2.397 & 34.87 \\
 5000 & 210.287 &  5.246 & 40.09 \\
10000 & 414.332 & 10.063 & 41.18 \\
\multicolumn{4}{c}{complex quad double arithmetic} \\
\#paths &  CPU  & GPU &  speedup \\ \hline
  10 &    5.859 &  2.696 &   2.17 \\
  20 &   11.189 &  2.852 &   3.92 \\
  50 &   24.018 &  2.866 &   8.38 \\
 100 &   38.782 &  2.966 &  13.08 \\
 200 &   67.703 &  3.568 &  18.97 \\
 500 &  174.769 &  6.203 &  28.17 \\
1000 &  368.449 & 11.175 &  32.97 \\
2000 &  851.255 & 21.432 &  39.72 \\
5000 & 2164.485 & 48.495 &  44.63 \\
\end{tabular}
\label{tabpathcyclic10}
\end{center}
\end{table}

\begin{figure}[hbt]
\begin{center}
\epsfig{figure=multevalcyclic10speedups.eps, width=9.5cm}
\epsfig{figure=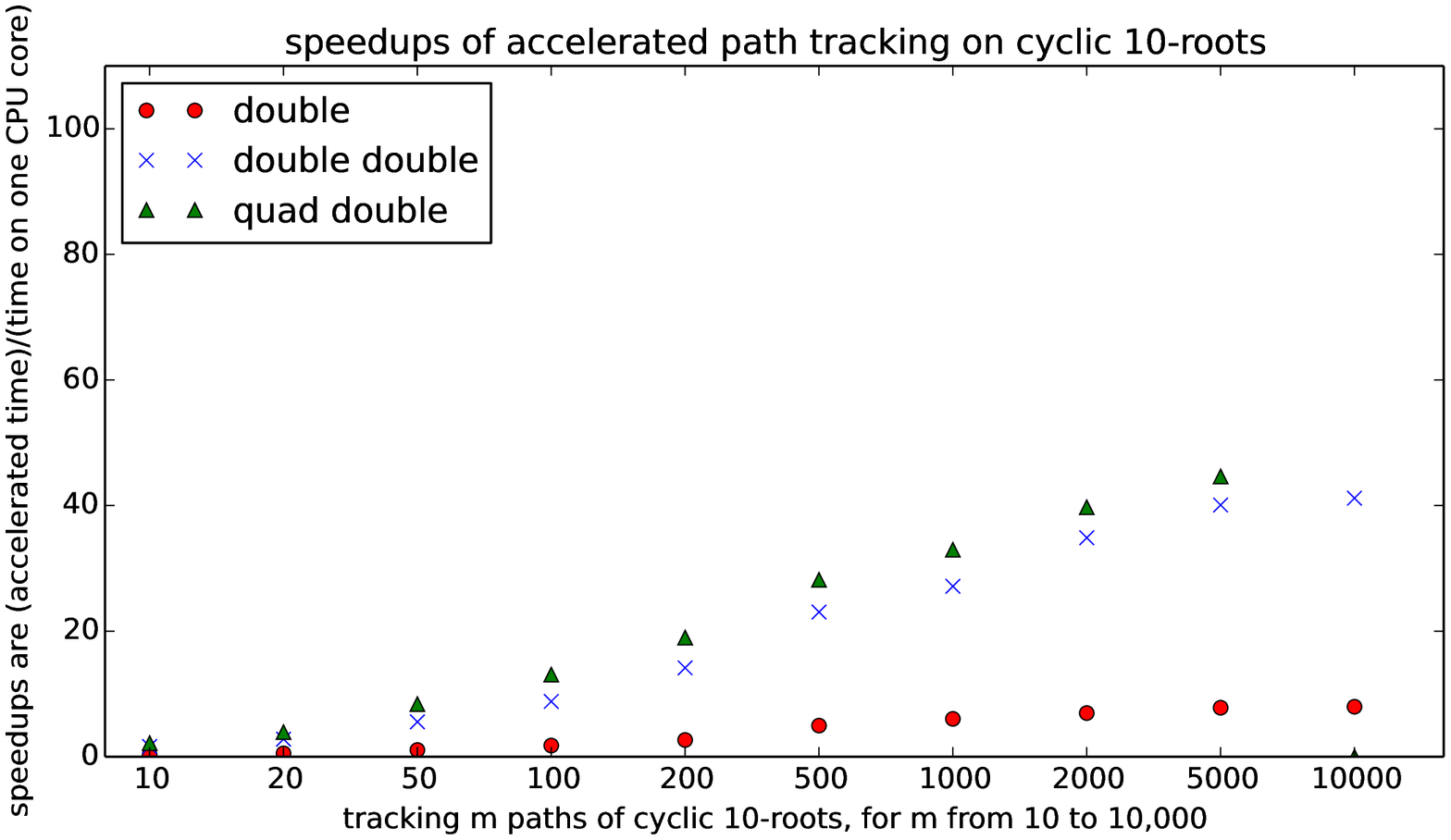, width=9.5cm}
\caption{The speedups for tracking many paths of the cyclic 10-roots
problem, compared to speedups for evaluating and differentiating,
in complex double, double double, and quad double arithmetic.}
\label{figpathcyclic10}
\end{center}
\end{figure}

The data in Table~\ref{tabpathcyclic10} is visualized
in Figure~\ref{figpathcyclic10}.
Compared to the speedups for evaluation and differentiation,
the speedups for path tracking are roughly about half of 
those of evaluation and differentiation, for double double
and quad double arithmetic.

\begin{figure}[hbt]
\begin{center}
\epsfig{figure=multpathcyclic10speedups.eps, width=9.5cm}
\epsfig{figure=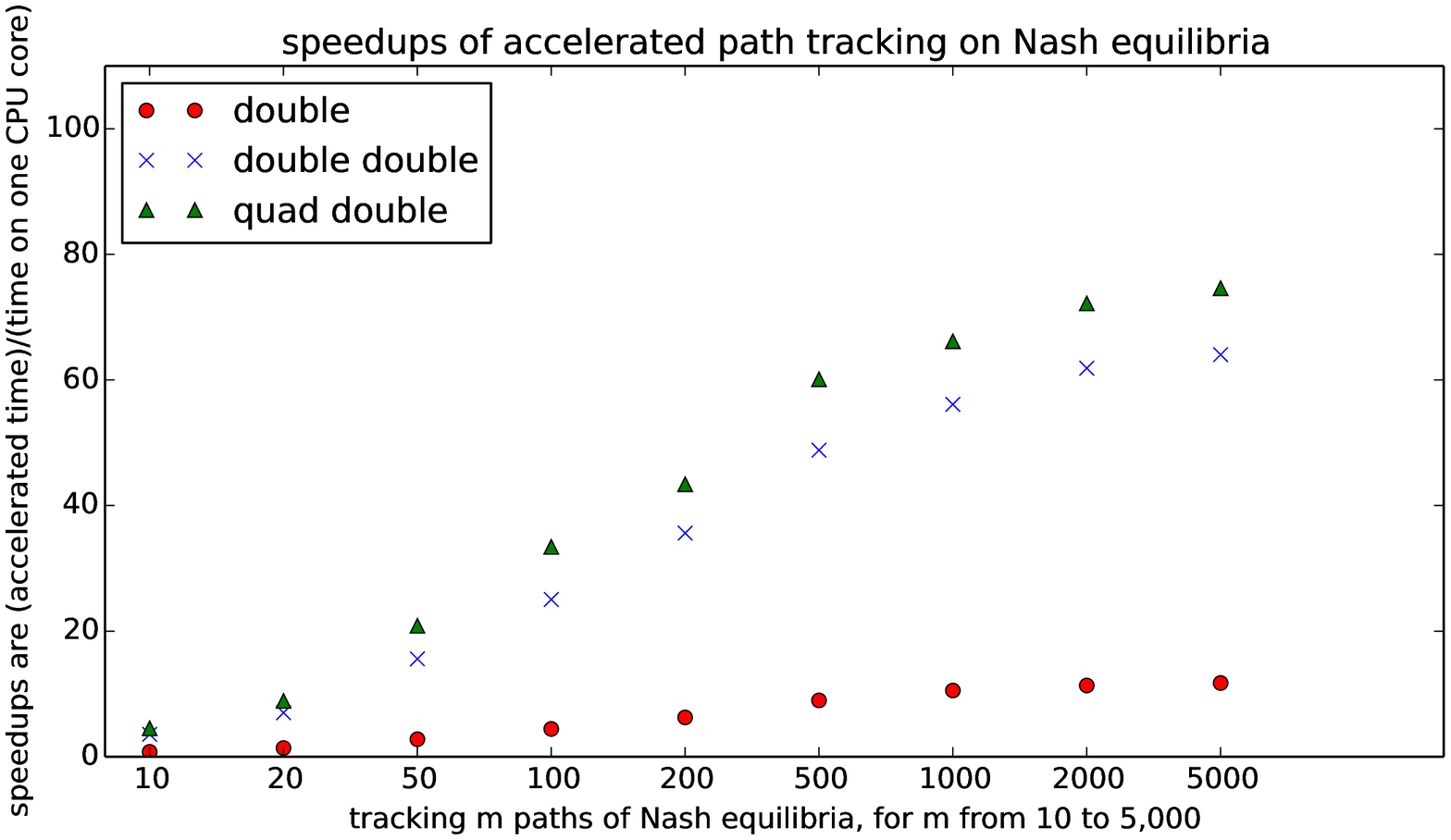, width=9.5cm}
\epsfig{figure=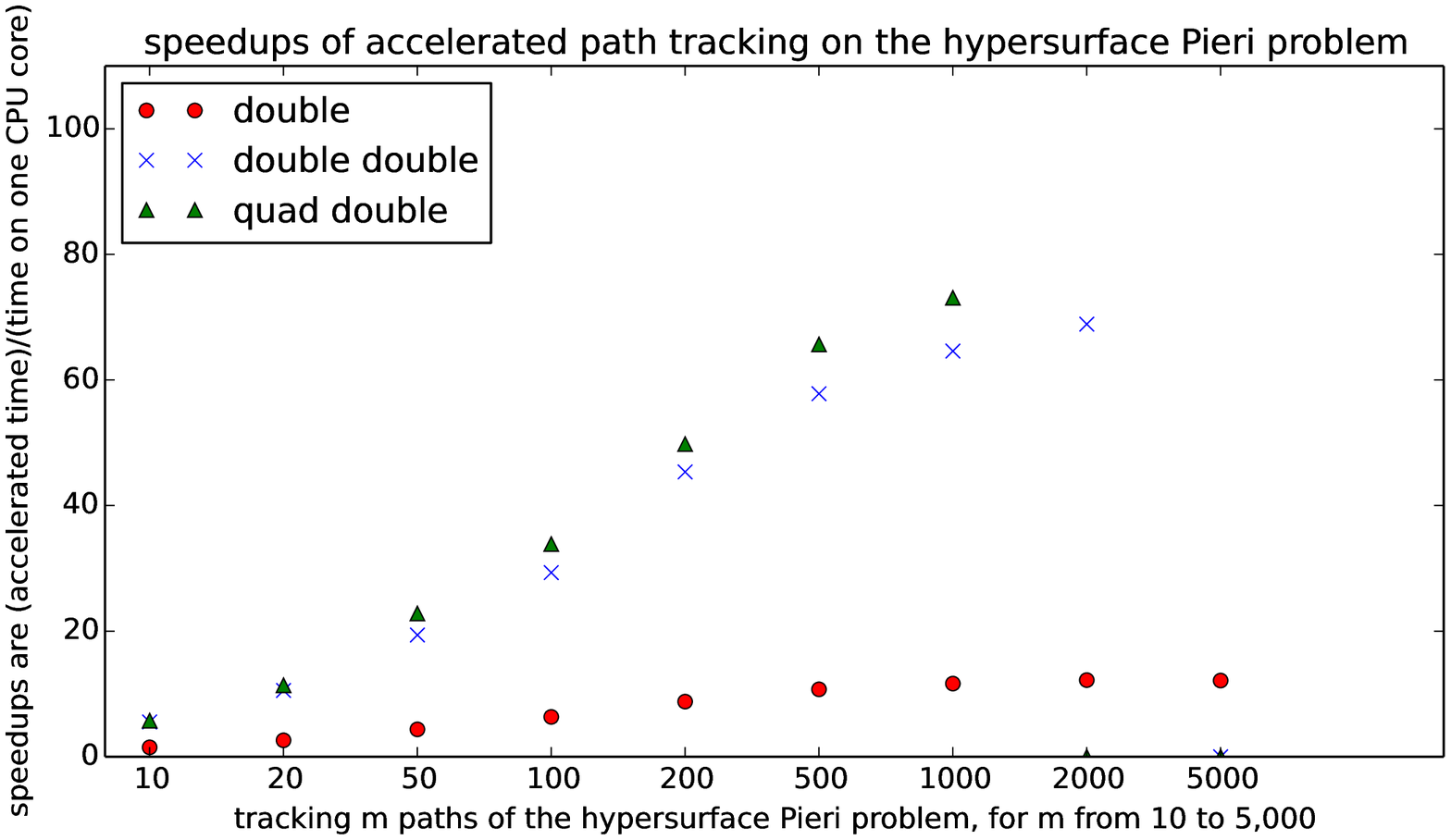, width=9.5cm}
\caption{The speedups for tracking many paths of the cyclic 10-roots
problem, compared to speedups for tracking many paths of the Nash
equilibrium system and of the hypersurface Pieri system,
in complex double, double double, and quad double arithmetic.}
\label{figpathnash8}
\end{center}
\end{figure}

Table~\ref{tabpathnash8} lists times and speedups for tracking
many paths of the Nash equilibrium system.
In Figure~\ref{figpathnash8} we visualize these data.
Notice that, as the Nash equilibrium system has more monomials
than the cyclic 10-roots system, the speedup for {\tt nash8}
are better than those form {\tt cyclic10}.
The speedups improve slightly for the Pieri problem,
but with a larger of number of monomials 
the memory allows for fewer paths to be tracked simultaneously.

%\bibliographystyle{plain}
%\bibliography{GPUphc}

% \appendix{Appendix A. times and speedups for {\tt nash8}}
% 
% Table~\ref{tabevalnash8} lists times and speedups for evaluating and
% differentiating the Nash equilibrium system.
% Times for path tracking are listed in Table~\ref{tabpathnash8}.

% \appendix{Appendix B. times and speedups for {\tt pieri44}}
% 
% Table~\ref{tabevalpieri44} lists times and speedups for evaluating and
% differentiating the hypersurface Pieri system.
% Times for path tracking are listed in Table~\ref{tabpathpieri44}.

\end{document}